%% file: main_v1.tex
\documentclass[pdflatex,sn-nature,oneside]{sn-jnl}

\usepackage{graphicx}
\usepackage{multirow}
\usepackage{amsmath,amssymb,amsfonts}
\usepackage{amsthm}
\usepackage{mathrsfs}
\usepackage[title]{appendix}
\usepackage{xcolor}
\usepackage{textcomp}
\usepackage{manyfoot}
\usepackage{booktabs}
\usepackage{algorithm}
\usepackage{algorithmicx}
\usepackage{algpseudocode}
\usepackage{listings}
\usepackage[labelformat=empty]{caption}
\usepackage{braket}
\usepackage{geometry}
\usepackage{fancyhdr}

\usepackage{times}
\usepackage{glossaries}
\usepackage{bm}
\usepackage[labelformat=empty]{caption}
\input{./glossary.tex}

\begin{document}
\vspace{-3cm}

\title[Article Title]{Beyond-Hubbard pairing in a cuprate ladder}


\author[1]{\fnm{Hari} \sur{Padma}}
\author[2,3]{\fnm{Jinu} \sur{Thomas}}
\author[1]{\fnm{Sophia} \sur{TenHuisen}}
\author[4]{\fnm{Wei} \sur{He}}
\author[1]{\fnm{Ziqiang} \sur{Guan}}
\author[5]{\fnm{Jiemin} \sur{Li}}
\author[6,7]{\fnm{Byungjune} \sur{Lee}}
\author[8]{\fnm{Yu} \sur{Wang}}
\author[8]{\fnm{Seng Huat} \sur{Lee}}
\author[8]{\fnm{Zhiqiang} \sur{Mao}}
\author[9]{\fnm{Hoyoung} \sur{Jang}}
\author[5]{\fnm{Valentina} \sur{Bisogni}}
\author[5]{\fnm{Jonathan} \sur{Pelliciari}}
\author[4]{\fnm{Mark P. M.} \sur{Dean}}
\author[2,3]{\fnm{Steven} \sur{Johnston}}
\author[1]{\fnm{Matteo} \sur{Mitrano}}

\affil[1]{\orgdiv{Department of Physics}, \orgname{Harvard University}, \orgaddress{\city{Cambridge}, \state{MA}, \country{USA}}}

\affil[2]{\orgdiv{Department of Physics \& Astronomy}, \orgname{University of Tennessee}, \orgaddress{\city{Knoxville}, \state{TN}, \country{USA}}}

\affil[3]{\orgdiv{Institute of Advanced Materials and Manufacturing}, \orgname{University of Tennessee}, \orgaddress{\city{Knoxville}, \state{TN}, \country{USA}}}

\affil[4]{\orgdiv{Condensed Matter Physics and Materials Science Department}, \orgname{Brookhaven National Laboratory}, \orgaddress{\city{Upton}, \state{NY}, \country{USA}}}

\affil[5]{\orgdiv{National Synchrotron Light Source II}, \orgname{Brookhaven National Laboratory}, \orgaddress{\city{Upton}, \state{NY}, \country{USA}}}

\affil[6]{\orgdiv{Department of Physics}, \orgname{Pohang University of Science and Technology}, \orgaddress{\city{Pohang}, \country{Korea}}}

\affil[7]{\orgdiv{Max Planck POSTECH/Korea Research Initiative}, \orgname{Center for Complex Phase Materials}, \orgaddress{\city{Pohang}, \country{Korea}}}

\affil[8]{\orgname{2D Crystal Consortium}, \orgname{Pennsylvania State University}, \orgaddress{\city{University Park}, \state{PA}, \country{USA}}}

\affil[9]{\orgdiv{PAL-XFEL}, \orgname{Pohang Accelerator Laboratory, POSTECH}, \orgaddress{\city{Pohang}, \country{Korea}}}

\baselineskip24pt %

\maketitle

\vspace{-8mm}

\textbf{The Hubbard model is believed to capture the essential physics of cuprate superconductors. However, recent theoretical studies suggest that it fails to reproduce a robust and homogeneous superconducting ground state. Here, using resonant inelastic x-ray scattering and density matrix renormalization group calculations, we show that magnetic excitations in the prototypical cuprate ladder Sr$_{14}$Cu$_{24}$O$_{41}$ are inconsistent with those of a simple Hubbard model. The magnetic response of hole carriers, contributing to an emergent branch of spin excitations, is strongly suppressed. This effect is the consequence of $d$-wave-like pairing, enhanced by nearly an order of magnitude through a large nearest-neighbor attractive interaction. The similarity between cuprate ladders and the two-dimensional compounds suggests that such an enhanced hole pairing may be a universal feature of superconducting cuprates.}

\vspace{2mm}

Understanding how electrons pair in strongly correlated materials such as the cuprate superconductors is a major unsolved problem in condensed matter physics \cite{norman2011challenge,scalapino2012common}.
While pairing in conventional superconductors occurs via electron-phonon interaction, the pairing mechanism and onset of superconductivity in the presence of strong electron-electron interactions is a correlated many-body problem that presents additional challenges. The single-band Hubbard model, where electrons in a lattice hop with a characteristic energy $t$ and interact through an on-site Coulomb repulsion $U$, is widely regarded as the minimal description of the cuprates \cite{scalapino2012common, Arovas2022hubbard}. This model successfully captures the presence of antiferromagnetism, charge order, a pseudogap, and a Fermi-liquid-like regime \cite{white1989numerical, Gull2010momentum, scalapino2012common, Huang2018stripe, qin2022hubbard}.
However, it is unclear whether the Hubbard model alone yields a robust $d$-wave superconducting ground state over competing spin- and charge-ordered phases \cite{Gull2013superconductivity, zheng2017stripe, qin2020absence}. Furthermore, recent calculations suggest that its low-energy antiferromagnetic spin fluctuations can only account for half of the total pairing interaction \cite{Dong2022quantifying}. Together with the pronounced sensitivity of superconductivity to band structure effects~\cite{jiang2019superconductivity} and additional interactions~\cite{Johnston2010systematic, jiang2022enhancing, zhang2022enhancement, peng2023enhanced, zhou2023robust}, these findings strongly suggest that the Hubbard model provides an incomplete description of cuprate superconductivity.

Since antiferromagnetic correlations play a crucial role in the behavior of the high-T$_c$ cuprates, a powerful strategy to shed light on missing many-body interactions is to probe their magnetic excitation spectrum and compare it with the predictions of different model Hamiltonians. Cuprate ladders (see Fig.~1(a)) are uniquely well-suited to this endeavor \cite{dagotto1999experiments, Vuletic2006spinladder}. Unlike two-dimensional copper oxides, cuprate ladders can be accurately described using modern \gls*{DMRG} methods and their magnetic excitations can be computed with excellent energy and momentum resolution. At the same time, these systems retain the key properties of the high-$T_{\mathrm{C}}$ superconducting compounds, including charge order \cite{abbamonte2004crystallization} and superconductivity \cite{uehara1996superconductivity, nagata1998pressure}. Further, the CuO$_2$ planes in presence of charge stripes in the 2D high-$T_\mathrm{C}$ cuprates have previously been modeled in terms of coupled ladders \cite{arrigoni2004mechanism,jiang2022stripe} with magnetic excitations that could explain the ``hourglass'' dispersion observed in underdoped compounds \cite{tranquada2004quantum, tranquada2020cuprate}, underscoring the similarity between ladder compounds and two-dimensional cuprates. 

The magnetic excitation spectrum of cuprate ladders is theoretically known to be sensitive to both the underlying electronic interactions and the presence of doped carriers. The ground state of an isotropic, undoped ladder consists of spin singlets \cite{dagotto1992superconductivity} and its elementary magnetic excitations are singlet-to-triplet transitions (``triplons,'' see Fig.~1(b)). Each doped hole breaks a spin singlet and behaves like a quasiparticle with loosely bound spin and charge components \cite{troyer1996properties, liu2016nature}, which contribute distinct spin flip excitations \cite{troyer1996properties, Kumar2019ladders}. As sketched in Fig.~1(c) (for $H=0$, see Appendix), the magnetic excitation spectrum of an undoped Hubbard ladder features a continuum of two-triplon excitations, whereas a doped ladder exhibits an additional lower-energy branch from quasiparticle spin flips. However, the experimental identification of magnetic excitations from doped carriers in cuprate ladders has been impeded by significant limitations in energy resolution and counting statistics.

Here, we report high-resolution ($\sim$35 meV) Cu $L$-edge \gls*{RIXS} measurements of the magnetic excitations in the cuprate ladder Sr$_{14}$Cu$_{24}$O$_{41}$. In this compound, the ladders are self-doped with $\sim 0.06$ holes/Cu atom \cite{nucker2000hole, osafune1997optical,Vuletic2006spinladder}, which form a charge ordered state \cite{abbamonte2004crystallization} below $T_{\mathrm{CO}}$ of 250 K (see SM Section 1). The presence of self-doped holes in the parent compound provides a unique opportunity to probe the magnetic excitations of the doped holes without the added complexity of chemical substitution. The raw \gls*{RIXS} spectra measured at 260 K (Fig.~2(a), see SM Section 2 for raw RIXS spectra at 40 K) along the leg direction show peaks corresponding to the elastic line, a 60-meV bond-stretching phonon, its second harmonic, and dispersive magnetic excitations with spectral weight up to 600 meV. We isolate the magnetic excitations by fitting and subtracting elastic and phonon contributions as shown in Fig.~2(b) (see SM Section S2 for fit details). We then normalize the subtracted spectra using the intensity of the $dd$ orbital excitations and the spin-flip scattering cross-section calculated using a single-ion model~\cite{MorettiSala2011energy,Wang2019EDRIXS}. This procedure yields the dynamical spin structure factors shown in Figs.~2(c) and 2(d) (see SM Section 2 for further details). The magnetic excitations are dominated by an intense two-triplon continuum, with a shape and dispersion consistent with previous experimental and theoretical results \cite{Schlappa2009collective, Tseng2022crossover, troyer1996properties, Kumar2019ladders}. 

We now discuss the key spectroscopic feature of this work. By leveraging our high energy resolution, we detect a previously unobserved dispersive magnetic peak below the two-triplon continuum, near the zone boundary (Figs.~2(b) and 2(d), see additional momenta in SM Section 2). Its dispersion closely follows the low-energy shoulder of the two-triplon continuum, consistent with the expected behavior of the quasiparticle spin flip branch \cite{troyer1996properties, Kumar2019ladders}. However, its intensity is strongly suppressed compared to theoretical expectations for a doped Hubbard ladder \cite{troyer1996properties, Kumar2019ladders}. The quasiparticle spin flip branch is completely absent at low temperature ($T<T_{\mathrm{CO}}$), and only faintly present in the high-temperature spectra ($T>T_{\mathrm{CO}}$), as shown in Figs.~2(b) and 2(d). While the absence of the quasiparticle spin flip branch at low temperature might be associated with the opening of the charge order gap, this cannot explain its weak intensity above $T_{\mathrm{CO}}$. Disorder alone, which can pin holes to the lattice and modify the spin flip spectral weight, may also be ruled out as a dominant factor. A previous study on the Ca-doped compound~\cite{Tseng2022crossover} showed that the disordered Hubbard model predicts a broadened and flattened dispersion of the two-triplon continuum. In contrast, the parent compound we consider here is not disordered by chemical substitution and exhibits a well-defined and strongly dispersive two-triplon continuum.

We investigate the origin of this intensity discrepancy by comparing the \gls*{RIXS} data with \gls*{DMRG} calculations of the dynamical spin structure factor of a single-band Hubbard ladder. We first extract the parameters of the undoped Hubbard ladder by fitting the experimental two-triplon dispersion at 40 K using a Bayesian optimization procedure (see Appendix). We obtain a nearest-neighbor leg hopping $t = 0.38$~eV, rung hopping $t_\perp = 0.84t$, a diagonal hopping $t^\prime = -0.3t$, and an on-site Coulomb repulsion $U = 8t$. The corresponding superexchange interactions are $J = -4t^2/U = 190$ meV and $J_\perp =-4t_\perp^2/U = 134$ meV, which are in excellent agreement with reported values extracted from inelastic neutron scattering \cite{notbohm2007one}. We then use these parameters to calculate the dynamical spin structure factor of the two-leg Hubbard ladder upon 6.25\% hole doping \cite{nucker2000hole, osafune1997optical} using the Krylov space correction vector method~\cite{DMRGKrylov} (see SM Section 3). As shown in Fig.~3(a), the \gls*{DMRG} spectra feature an intense branch of quasiparticles spin flips with a downward dispersion at the Brillouin zone boundary. This result is consistent with prior calculations~\cite{Kumar2019ladders}, but is in stark contrast with the measured magnetic spectra. We note that the three-band Hubbard model also cannot account for our experimental data. Prior work shows that the doped low-energy magnetic excitation spectra of the three-band model are nearly identical to their effective single-band counterparts~\cite{Li2021particle}.

Such a disagreement suggests the presence of additional interactions suppressing the free propagation of spin-1/2 quasiparticles that would otherwise disrupt the singlet background. Since the simple Hubbard model assumes only an onsite Coulomb repulsion, which might not sufficiently account for nonlocal interactions, we consider the effect of an additional nearest-neighbor Coulomb interaction $V$. A repulsive interaction ($V > 0$) does not significantly alter the calculated spin structure factor (see SM Fig. S8). However, the introduction of a nearest-neighbor attraction ($V < 0$) dramatically suppresses the intensity of the quasiparticle spin flip excitations and sharpens the two-triplon continuum. We obtain the best agreement between experiment and \gls*{DMRG} calculations by introducing an attractive interaction $V$ of order $-1.0t$ to $-1.25t$ (see Fig.~3(b-c)). A closer inspection of the \gls*{DMRG} spectra at selected momenta, presented in Fig.~3(d-e), reveals that the suppression of magnetic excitations from the doped holes is monotonic with increasing $V$. An ARPES study in the different context of holon states in the chain compound Ba$_{2-x}$Sr$_x$CuO$_{3+\delta}$ obtained a very similar value of $V$, in the range $-0.8t$ to $-1.2t$ \cite{chen2021anomalously}, supporting the idea that this interaction is robust and shared between different cuprate families. Such an attractive interaction has been posited to arise from electron-phonon coupling~\cite{chen2021anomalously, wang2021phonon}, which is known to be significant in two-dimensional cuprates~\cite{lanzara2001evidence,he2018rapid} as well as cuprate ladders \cite{Adamus2023analogies}. Alternatively, an effective attractive interaction between doped holes could also emerge from the non-uniform electric polarizability of the system~\cite{derriche2024non}.

Next, we examine the effect of this nearest-neighbor attraction on the magnetic excitations by analyzing its influence on holes. Fig.~3(d) illustrates that in an isotropic Hubbard ladder (where $V=0$), doped holes disrupt rung singlets. These holes are weakly paired, with a correlation length of two to three lattice constants for model parameters consistent with Sr$_{14}$Cu$_{24}$O$_{41}$~\cite{dagotto1992superconductivity, troyer1996properties}. Introducing a negative $V$ makes it energetically favorable for holes to form tightly bound pairs on the same rung. Consequently, more unbroken spin singlets are available for excitation into triplets, enhancing the intensity of the two-triplon excitations at the cost of the quasiparticle spin flip continuum. That is, the suppression of the quasiparticle spin-flip branch is a direct consequence of enhanced hole pairing. Our DMRG calculations show that the attractive $V$ enhances the hole-hole binding energy by almost an order of magnitude relative to the pure Hubbard model, with the value increasing to 0.074$t$ for $V = -1.25t$ (see Fig.~3(f)). 

Based on the intensity suppression observed above $T_{\mathrm{CO}}$, we underscore that the enhanced hole pairing is present at temperatures above the onset of charge order and is not a consequence of the formation of a charge order gap. Importantly, while we illustrate hole pairs along the rungs for simplicity in Fig.~3(e)), the attractive $V$ in our model is isotropic along the leg and rung directions. Our calculations (see SM Fig. S9) show that the resulting pair-pair correlations for such an interaction retain the same quasi-$d$-wave-like structure as obtained for $V = 0$, where the pair correlations along the rung are opposite in sign to those along the leg despite the lack of local $C_4$ symmetry about each site. This conclusion is consistent with another recent study that examined a two-leg ladder with isotropic hopping~\cite{Dolfi2015pair}. 

Finally, we investigate the effective dimensionality of the ladder, and consider whether our observation of enhanced pairing due to a large attractive Coulomb interaction can be generalized to other cuprate families. In chain cuprates, the fundamental spin, charge, and orbital degrees of freedom separate into collective excitations propagating with different velocities, whereas these collective modes are confined in higher dimensions~\cite{Lake2010confinement}. In Fig.~4, we present evidence of spin-orbital separation in the \gls*{RIXS} spectra. The excitation of Cu 3d $x^2 - y^2$ holes into different orbital states gives rise to a manifold of transitions between 1.6 and 2.7 eV. We focus on the $xz/yz$ feature, which displays a characteristic orbiton dispersion similar to that observed in cuprate chains \cite{schlappa2012spin, wohlfeld2013microscopic, wohlfeld2019propagation}. The lower branch of the $xz/yz$ orbiton in Fig.~4(b) is well described by the Kugel-Khomskii model \cite{kugel1982jahn} (refer to SM Section 2 for details). We obtain an effective orbital superexchange parameter $J_O = 22.5 \pm 5.4$ meV, which quantifies the propagation of orbital excitations along the ladder legs. This value is substantially lower than $J_O \sim 75$ meV found in Sr$_2$CuO$_3$ \cite{schlappa2012spin} and indicates a higher degree of orbiton confinement when compared to cuprate chains. Furthermore, the spin-orbital separation in Sr$_{14}$Cu$_{24}$O$_{41}$ is not accompanied by spin-charge separation \cite{troyer1996properties, liu2016nature}, as evidenced by the gapped two-triplon continuum (see Fig.~2), which significantly departs from the gapless spinon dispersion observed in chain compounds \cite{schlappa2012spin} and extremely anisotropic ladders \cite{Bisogni2015orbital}. These observations place  Sr$_{14}$Cu$_{24}$O$_{41}$ in a cross-over regime between one- and two-dimensions and indicate that our findings may be directly relevant to the isotropic two-dimensional CuO$_2$ plane.

Our observation of enhanced hole pairing due to an attractive Coulomb interaction in a cuprate ladder has broader implications for the understanding of how superconductivity is stabilized in the copper oxides. On the one hand, our data indicates that an attractive intersite interaction may be a crucial addition to the minimal Hubbard model for cuprates beyond spin chain compounds. On the other hand, the large binding energy and the $d$-wave symmetry of the hole pairing, together with recent work on related theoretical models ~\cite{jiang2022enhancing, peng2023enhanced}, suggest that an attractive $V$ might be important in enhancing superconducting correlations over charge order and other symmetry breaking phenomena that dominate in a pure Hubbard model. Our work provides an experimental basis for what may be the key missing ingredient in the theoretical description of robust $d$-wave superconductivity in two-dimensional hole-doped cuprates and motivates the search for a similar minimal model in the electron-doped sector of the cuprate phase diagram. 

\clearpage

\begin{figure}[h] 
    \centering
    \includegraphics[width=0.95\textwidth]{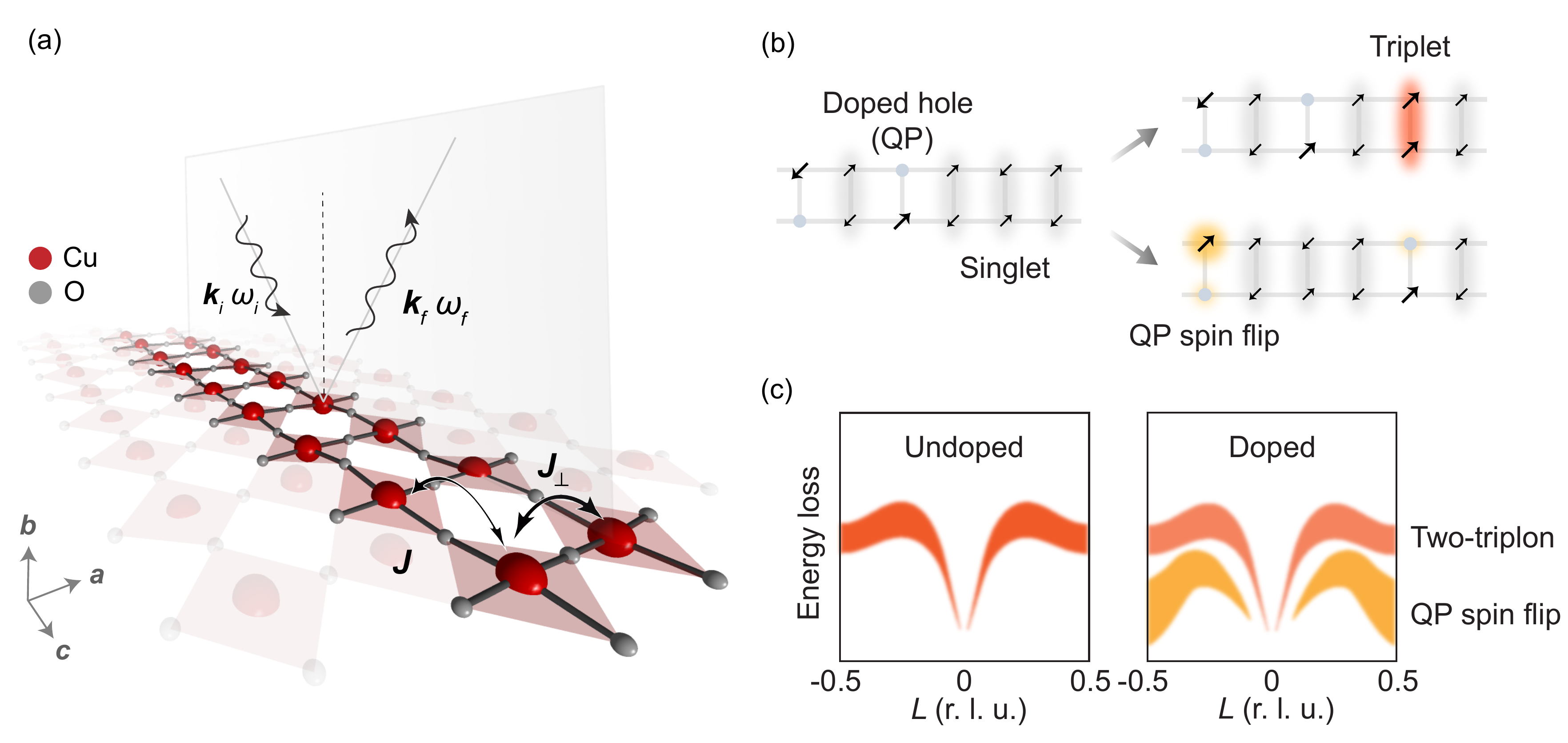} 
    \caption{ { FIG.~1. Magnetic excitations of the cuprate ladder Sr$_{14}$Cu$_{24}$O$_{41}$.} (a) Sketch of resonant inelastic X-ray scattering off the cuprate ladder compound Sr$_{14}$Cu$_{24}$O$_{41}$. The ladder legs and rungs run parallel to the $c$ and $a$ axes, respectively. The scattering plane is spanned by the sample $b$ and $c$ axes. $J$ and $J_\perp$ indicate the exchange couplings along the ‘leg’ and ‘rung’ directions, respectively. $\bm{k}_i$ ($\bm{k}_f$) and $\omega_i$ ($\omega_f$) denote initial (final) momentum and energy of the X-ray photons. (b) A hole-doped spin ladder features loosely bound quasiparticles (QP) in a background of spin singlets. There are two types of magnetic excitations with $\Delta S=1$: singlet-to-triplet excitations, and quasiparticle spin flips. Black arrows and grey/yellow circles represent spins and holes, while grey and red shadings correspond to spin singlets and triplets, respectively. Singlet correlations along the legs are omitted for simplicity. (c) Expected dispersions of two-triplon (orange) and QP spin flip excitations (yellow) along the leg direction in an undoped (left) and hole-doped (right) ladder (for $H = 0$). Momenta are labeled in reciprocal lattice units (r. l. u.).}
    \label{fig:fig1}
\end{figure}

\clearpage

\begin{figure}[h] 
    \centering
    \includegraphics[width=\textwidth]{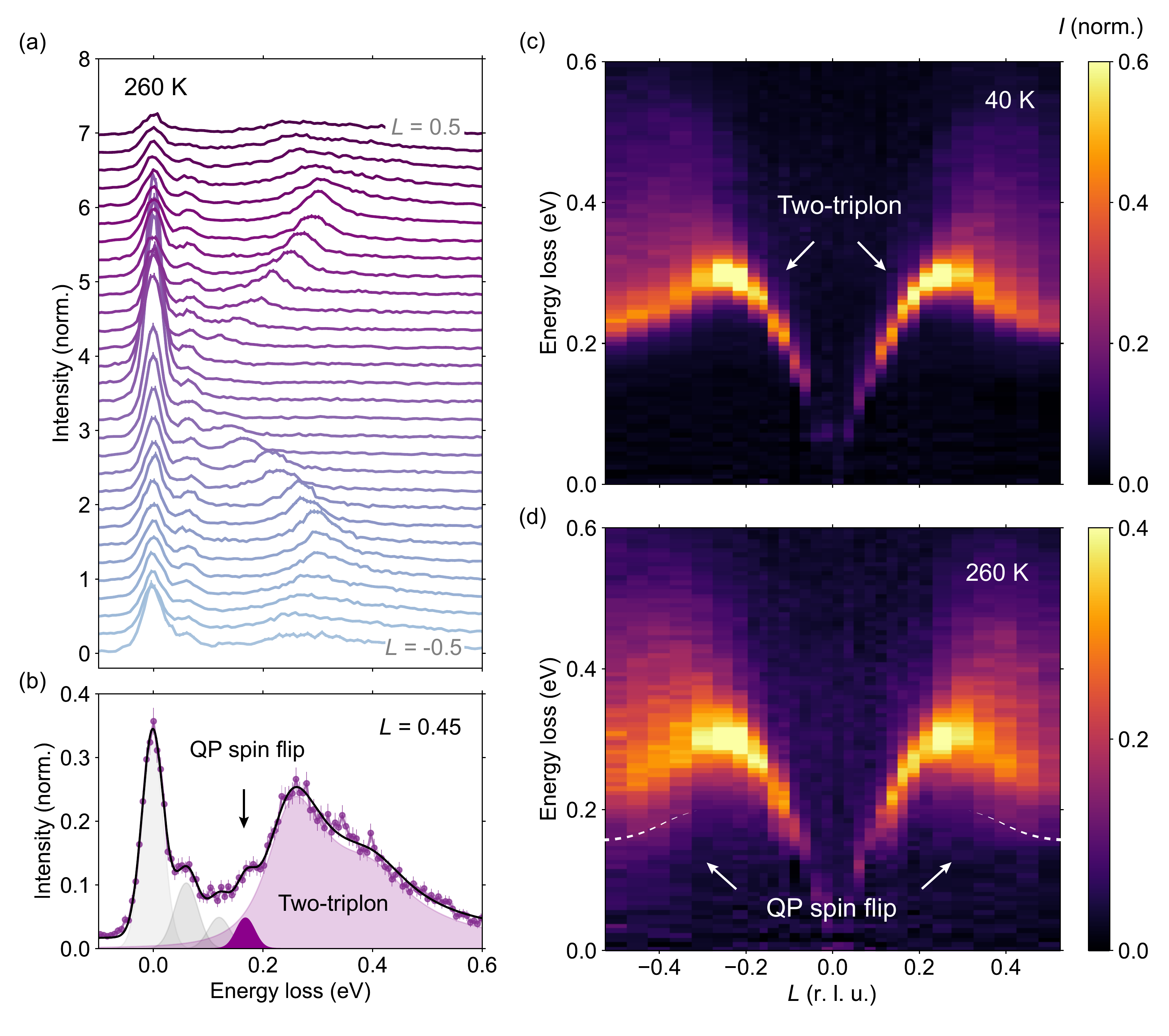} 
    \caption{{FIG.~2. Magnetic excitations from the doped holes.} (a) \gls*{RIXS} spectra along the leg direction for momenta spanning $L$ = -0.5 to 0.5 r.~l.~u., measured at 260 K. Spectra are vertically offset for clarity. (b) Representative fit to the low-energy region of the \gls*{RIXS} spectrum ($L$ = 0.45 r.~l.~u.) at 260 K (solid line), with the elastic peak, bond-stretching phonon, and its second harmonic fit to Gaussians, shaded in light grey. The quasiparticle spin flip feature is fit to a Gaussian, shaded in dark pink. The two-triplon continuum is fit to a phenomenological lineshape consisting of an asymmetric Lorentzian and a Gaussian, shaded in light pink. (c-d) Intensity map of the dynamical spin structure factor at 40 K and 260 K (below and above the charge order transition, respectively) as function of momentum and energy loss. We subtract elastic and phonon contributions, shown in (b), and normalize the data to the orbital excitation and spin-flip scattering cross-section as described in the SI. While the map at 40 K exhibits negligible spectral weight below the two-triplon continuum, the map at 260 K features an additional weak branch corresponding to quasiparticle spin flip excitations. The dashed white lines are a guide to the eye. The energy resolution of our \gls*{RIXS} measurement is $\sim$35 meV.}
    \label{fig:fig2}
\end{figure}

\clearpage

\begin{figure}[h] 
    \centering
    \includegraphics[width=0.9\textwidth]{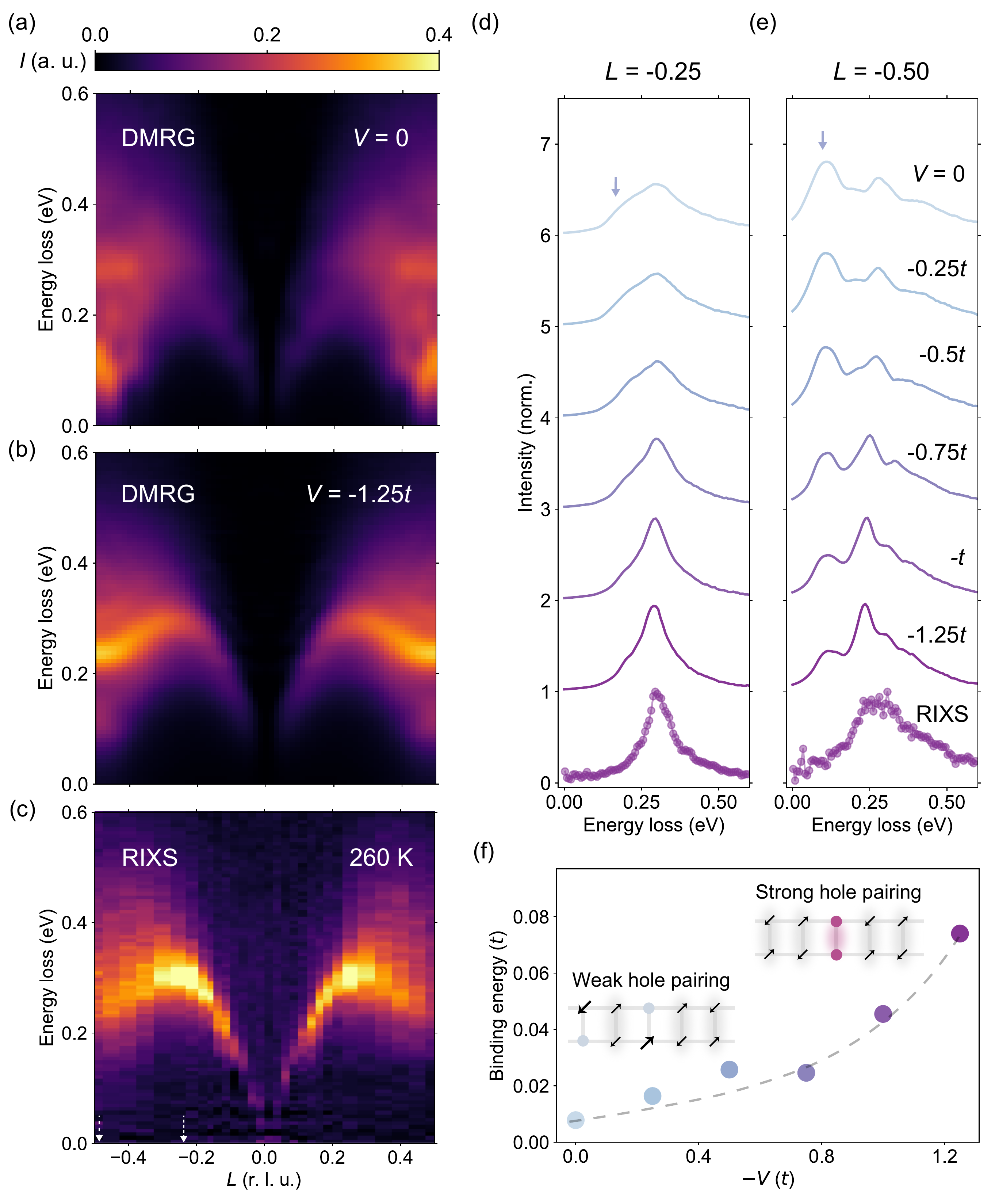} 
    \caption{{FIG.~3. Signatures of enhanced hole pairing due to an attractive nearest-neighbor interaction.} (a-b), Theoretical dynamical spin structure factors for $V = 0$ (a) and $-1.25t$ (b), calculated using \gls*{DMRG} on ladder clusters. (c) Experimental dynamical spin structure factor. White arrows indicate selected momenta shown in (d, e). (d-e) Experimental (filled circles) and theoretical (solid lines) dynamical spin structure factors at $L$ = -0.25 and -0.50 r. l. u. \gls*{DMRG} curves for varying $V$ are vertically offset for clarity. The best agreement between theory and experiment occurs for $V = -1.25t$. The pale blue arrows point to the quasiparticle spin-flip excitations. (f) The hole pair binding energy as a function of $V$. As $-V$ increases, holes tend to pair on neighboring sites. Arrows and circles represent spins and holes, respectively, while the purple shading represents the hole pair binding. The dashed line is a guide to the eye.}
    \label{fig:fig3}
\end{figure}

\clearpage

\begin{figure}[h] 
    \centering
    \includegraphics[width=1\textwidth]{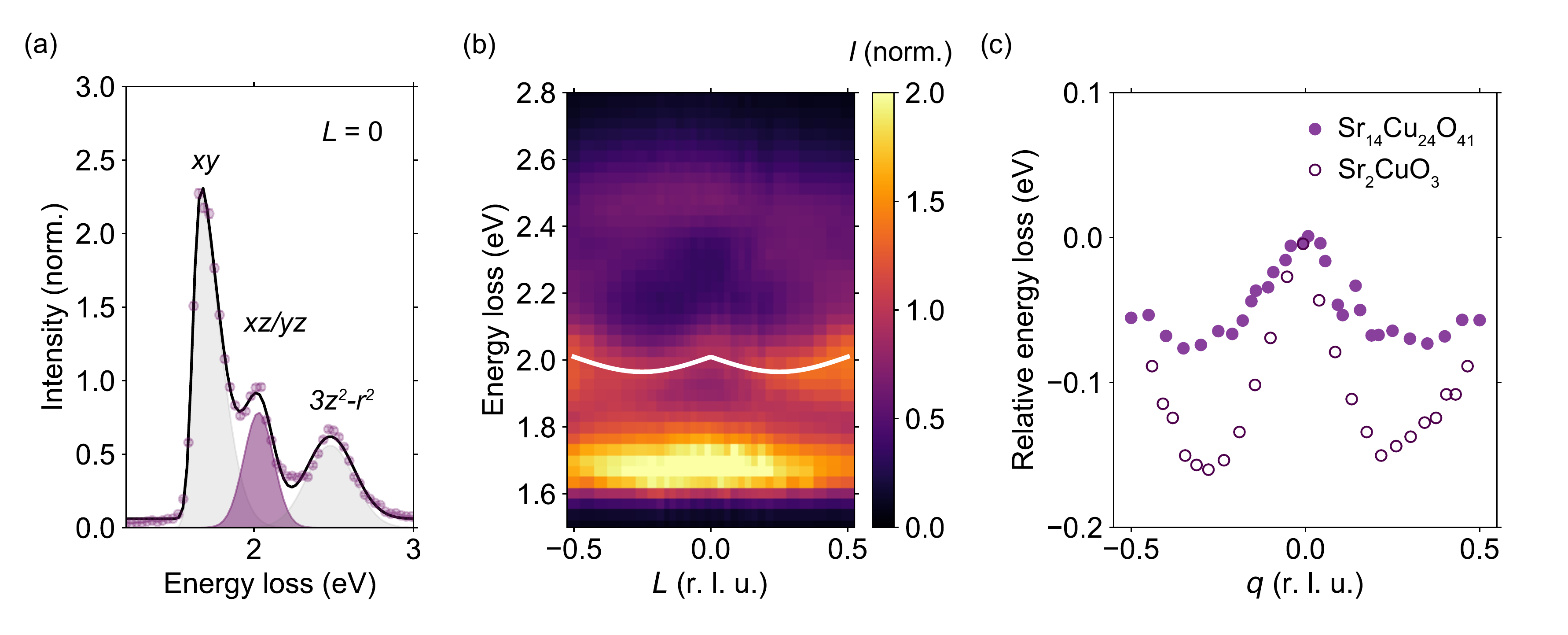} 
    \caption{{FIG.~4. Orbiton dispersion and dimensional crossover} (a), \gls*{RIXS} spectrum (circles) and fit (solid line) of the $dd$ orbital excitations at $L=0$. The fit includes three Lorentzian corresponding to excitations from the 3d$_{x^2 – y^2}$ orbital to $xy$ (grey), $xz$/$yz$ (purple), and 3$z^2-r^2$ (grey) orbitals. Note that here, $x$, $y$, and $z$ are parallel to the crystallographic $a$, $c$, and $b$ directions, respectively. (b), \gls*{RIXS} intensity map of the $dd$ excitations as function of momentum and energy loss. The white solid line is a fit to the lower edge of the orbiton dispersion derived from the Kugel-Khomskii Hamiltonian with $J_O=(22.5\pm5.4)$ meV. (c), Orbiton dispersion in cuprates with varying dimensionality. The ladder data are a fit to the spectra shown in (b), while the data for the cuprate chain, Sr$_2$CuO$_3$, are reproduced from \cite{schlappa2012spin}.}
    \label{fig:fig4}
\end{figure}

\clearpage

\section*{Data availability}\vspace{-4mm}
The data that support the findings of this study are present in the paper and/or in the Supplementary Material. Additional data related to the paper are available from the corresponding authors upon reasonable request.

\section*{Acknowledgements}\vspace{-4mm}
We thank P. Abbamonte, N. P. Armitage, C. D. Batista, E. Demler, M. Dressel, A. Georges, T. Giamarchi, S. A. Kivelson, A. J. Millis, S. Sachdev, M. Sentef, D. A. Tennant, and J. Tranquada for insightful discussions. This work was primarily supported by the U.S.\ Department of Energy, Office of Basic Energy Sciences, Early Career Award Program, under Award No.\ DE-SC0022883. The work by J.T.\ and S.J.\ (theory modeling) was supported by the U.S.\ Department of Energy, Office of Science, Office of Basic Energy Sciences, under Award Number DE-SC0022311. Work performed at Brookhaven National Laboratory (RIXS and RIXS data interpretation) was supported by the U.S.\ Department of Energy (DOE), Division of Materials Science, under Contract No.~DE-SC0012704. B.L. and H.J. were supported by the National Research Foundation of Korea (MSIT), Grant No. 2022M3H4A1A04074153 and 2020M3H4A2084417. This research used resources at the SIX beam line of the National Synchrotron Light Source II, a U.S. DOE Office of Science User Facility operated for the DOE Office of Science by Brookhaven National Laboratory under Contract No. DE-SC0012704. The work at the PAL-XFEL was performed at the RSXS endstation (Proposal No. 2023-1st-SSS-002), funded by the Korea government (MSIT). The single crystal growth work was performed at the Pennsylvania State University Two-Dimensional Crystal Consortium–Materials Innovation Platform (2DCC-MIP), which is supported by NSF Cooperative Agreement No.~DMR-2039351.

\section*{Author Contributions}\vspace{-4mm}
H.P., M.P.M.D., and M.M.\ conceived the experiment. H.P., S.T., W.H., Z.G., J.L., J.P., and V.B.\ conducted the \gls*{RIXS} measurements. J.T.\ and S.J.\ performed the \gls*{DMRG} calculations. H.P., S.T., B.L., and H.J.\ conducted the x-ray absorption spectroscopy and resonant soft x-ray diffraction measurements. Y.W., S.H.L., and Z.M. synthesized the single crystal samples. H.P. analyzed the data with assistance from M.P.M.D. H.P., M.P.M.D, S.J., and M.M.\ wrote the manuscript with input from all authors. M.M.\ and S.J.\ supervised the experimental and theoretical aspects of the project, respectively.


\clearpage

\section*{Appendix: Methods}

\subsection*{Sample growth and characterization}\vspace{-2mm}

High-quality single crystal samples of Sr$_{14}$Cu$_{24}$O$_{41}$ were grown using a modified traveling solvent floating zone (TSFZ) technique. We first synthesized pure polycrystalline Sr$_{14}$Cu$_{24}$O$_{41}$ powder via a solid-state reaction, which we used as a feed material rod. We then chose CuO as the flux during the TSFZ growth. The flux rod (seed) was prepared by mixing the Sr$_{14}$Cu$_{24}$O$_{41}$ and CuO powders with the mass ratio of Sr$_{14}$Cu$_{24}$O$_{41}$:CuO = 1:0.0163. 
The feed speed of the feed rod was tuned within the 0.81-2.2 mm/hour range to maintain stable growth, and the growth speed was 0.81 mm/hour. Both feed and seed rods were rotated in mutually opposite directions at 30 RPM. We obtained a crystal rod with a length of 2 cm and a diameter of 4 mm. X-ray diffraction measurements confirmed good crystallinity and a structure in agreement with previous reports~\cite{vanishri2009crystal}. Characterization of the hole doping and charge order are provided in the Supplementary Text (see Fig.~S1 and S2). 

\subsection*{RIXS measurements}\vspace{-2mm}

We conducted high-resolution \gls*{RIXS} measurements at the beamline 2-ID of the National Synchrotron Light Source II, Brookhaven National Laboratory. The incident X-rays were $\pi$-polarized and tuned at resonance with the Cu $L_3$-edge peak (931 eV) with a combined energy resolution of 35 meV. We oriented a single crystal sample using Laue diffraction, which was then cleaved in-situ and mounted with the $b$ and $c$ axes in the scattering plane for all measurements described in the main text, as shown in Fig.~1a. The lattice parameters are $a$ = 11.47 \AA, $b$ = 13.35 \AA, and $c$ =7$c_L$ = 10$c_C$ = 27.46 \AA, where the subscripts '$L$' and '$C$' represent the ladder and chain subunits, respectively \cite{etrillard2004structural} (see SM Section 1 for more details on crystal structure). Our measurements are performed with the scattering angle fixed at $150^\circ$, while varying the incident angle $\theta$ by rotating the sample about the $a$ axis from $12.9^\circ$ to $137.1^\circ$, corresponding to momentum transfers of $L = -0.5$ to $0.5$ r.\ l.\ u.\ (defined in units of $2\pi/c_L$), with $H =0$. Given the layered crystal structure of Sr$_{14}$Cu$_{24}$O$_{41}$, we neglect the dispersion along the $K$ direction. 

The symmetry of the ladder geometry implies that sectors of even and odd triplon number do not mix, due to their different parity with respect to reflection about the centerline of the ladder \cite{notbohm2007one}. Hence, one- and two-triplon contributions can be studied in isolation by measuring at $H = 0.5$ and $H = 0$, respectively. In particular, since $H$ is fixed to zero, all measurements reported in the main text are sensitive to excitations with even triplon number, which is dominated by the two-triplon continuum. The raw \gls*{RIXS} spectra collected at 40~K and 260~K are shown in Fig.~S3 and S4, respectively. We conducted an additional set of measurements with the $a$ and $b$ axes in the scattering plane to detect the one-triplon excitation. We collected spectra at $\theta = 74.6^\circ$ and $10.9^\circ$ corresponding to momenta of $H$ = 0 and -0.5 (with $L$ = 0). The spectra, presented in Fig.~S6, show a one-triplon peak at an energy loss of 210 meV, consistent with previous inelastic neutron scattering measurements \cite{notbohm2007one}.

\subsection*{DMRG calculations}

\subsubsection*{Model parameters}\vspace{-4mm}

We simulate an extended Hubbard model on a two-leg ladder. The model Hamiltonian is
\begin{align}
    H = &- t \sum_{i,l,\sigma}  (\hat{c}^\dagger_{i,l,\sigma} \hat{c}^{\phantom{\dagger}}_{i+1,l,\sigma}  + \text{h.c.}) \nonumber - \ t_{\perp} \sum_{i,\sigma}  (\hat{c}^\dagger_{i,1,\sigma} \hat{c}^{\phantom{\dagger}}_{i,2,\sigma}  + \text{h.c.}) \\ \nonumber
    &- t^\prime \sum_{i,\sigma}  (\hat{c}^\dagger_{i,1,\sigma} \hat{c}^{\phantom{\dagger}}_{i+1,2,\sigma} + \hat{c}^\dagger_{i,2,\sigma} \hat{c}^{\phantom{\dagger}}_{i+1,1,\sigma}  + \text{h.c.}) \\ 
    &+ U \sum_{i,l} \hat{n}_{i,l,\uparrow} \hat{n}_{i,l,\downarrow}  
    + V \sum_{\substack{\langle i,l,i^\prime,l^\prime\rangle \\ 
    \sigma,\sigma^\prime}}  \hat{n}_{i,l,\sigma} \hat{n}_{i^\prime,l^\prime,\sigma^\prime}, 
\end{align}
where $l=1,2$ indexes the ladder leg, and the index $i$ runs over sites along the leg of length $L$. $\hat{c}^\dagger_{i,l,\sigma} (\hat{c}^{\phantom{\dagger}}_{i,l,\sigma} )$  creates (annihilates) an electron at site $i,l$  with spin $\sigma$. $\hat{n}_{i,l,\uparrow} (\hat{n}_{i,l,\downarrow})$ is the number of up (down) electrons, $\langle \dots\rangle$ denotes a sum over nearest-neighbor sites along both the leg and rung directions, $t$ and $t_\perp$ are the nearest neighbor hopping parameters along the leg and rung directions, respectively, $t^\prime$ is the next-nearest neighbor hopping along the diagonal direction of the ladder unit, $U$ is the onsite Hubbard repulsion, and $V$ is the nearest-neighbor Coulomb interaction. 

Setting $t=0.38$ eV and $U=8t$, we tune for $t_\perp$ and $t^\prime$ by comparing the experimental two-triplon spectra at 40 K to calculations of $S(\bm{q},\omega)$ for undoped Hubbard ladders. $V$ is only introduced later for the doped ladder calculations. Since the overall bandwidth of the two-triplon dispersion is controlled by exchange interaction $J= -4t^2/(U-V)$ along the leg, we adjusted the value of $U$ once $V$ is introduced to keep the value of $J$ fixed. 

For the ladder system, the dynamical spin structure factor is defined as 
\begin{align}\label{eq:Sqw}
    S(\bm{q},\omega)&= \sum_{f,\sigma} | \langle \psi_f | \hat{S}^z_{\bm{q}} | \psi_0 \rangle |^2 \delta(E_f - E_0 + \omega), 
\end{align}
where $S^z_{{\bf q}} = \frac{1}{\sqrt{2L}}\sum_{i,l} e^{-\mathrm{i}{\bf q}\cdot {\bf r}_{i,l}}S^\mathrm{z}_{i,l}$ is the Fourier transform of the $z$-component of the local spin operator, $\bm{q}$ and $\omega$ are the net momentum and energy transfer into the system, $\psi_0$ is the ground state with energy $E_0$ and $f$ indexes all final states $\psi_f$ with energy $E_f$. To compute $S({\bf q},\omega)$, we work in real space and evaluate 
\begin{equation}\label{eq:Sij}
S_{c,j}(\omega) = -\frac{1}{\pi}\mathrm{Im}
\bra{\Psi_0}
\hat{S}_j^z \frac{1}{\omega - H + E_0+\mathrm{i}\eta} \hat{S}_c^z
\ket{\Psi_0}
\end{equation}
using the Krylov space correction vector method~\cite{DMRGKrylov}. We then perform a Fourier transform to obtain the dynamical structure factor in momentum space, as described in Ref.~\cite{Nocera2018doping}. Here, $c$ denotes the site in the middle of the cluster, and the operator $S_j^z$ measures the total $z$-component of the spin on site $j$. The sum of the $S_{c,j}(\omega)$ contributions from sites on the same rung corresponds to a momentum transfer of $q_\perp = 0$ along the rung.


All numerical simulations are computed using the \gls*{DMRG} method~\cite{White1992density, White1993density}, as implemented in the \gls*{DMRG}++ code~\cite{re:alvarez0209}. For all \gls*{DMRG} simulations used in \gls*{BO}, we use $16 \times 2$ clusters and keep up to $m=2000$ states in both ground state and dynamical runs. The broadening parameter $\eta=0.05t$ is set to match the experimental resolution. Calculations for $S({\bf q},\omega)$ were performed on longer $64\times 2$ ladders and also kept $m = 2000$ states with a broadening of $\eta=0.1t$. 

We extracted the values of $t_\perp$ and $t^\prime$ from the \gls*{RIXS} data using the \gls*{BO} technique, a machine learning approach to find the global extrema of functions whose form is otherwise unknown~\cite{snoek2012practical}. Since the $S(\bm{q},\omega)$ depends non-trivially on the underlying Hamiltonian, \gls*{BO} is uniquely suited to tune for its parameters provided we define an appropriate cost function for the minimization. Here we choose
\begin{equation}
    C = L_2 \left[ S_\text{expt.}(\bm{q},\omega) - S^\prime_\text{theory}(\bm{q},\omega) \right],
\end{equation} where $L_2$ is the square root of the sum of the squared vector values obtained, and $S^\prime(\bm{q},\omega)_\text{theory}$ is the \gls*{DMRG} spectra interpolated onto the experimental grid. We only use the spectra with $\bm{q}>0$ in our analysis. To apply \gls*{BO}, we use the \textit{suggest} utility in the BayesianOptimization~\cite{BOpackage} package. For each value of the cost function we report for a particular set of parameters, the optimizer suggests a new set. We iterate this process until the suggestions converge. The best-fit values are given in the manuscript. 

\subsubsection*{Binding energy}\vspace{-3mm}

We compute the binding energy ($B.E.$) for the hole doped $64 \times 2$ ladders using 
\begin{equation}
    B.E. = 2  E_{GS}(N-1) - E_{GS}(N) - E_{GS}(N-2),
\end{equation}
where $E_{GS}$ is the ground state energy and $N=120$ is the number of electrons. We keep all other parameters unchanged. Per our definition, $B.E. > 0$ indicates hole-pairing.  

\clearpage

\bibliography{references}

\end{document}


\title{Beyond-Hubbard pairing in a cuprate ladder} 

\author[1]{\fnm{Hari} \sur{Padma}}
\author[2,3]{\fnm{Jinu} \sur{Thomas}}
\author[1]{\fnm{Sophia} \sur{TenHuisen}}
\author[4]{\fnm{Wei} \sur{He}}
\author[1]{\fnm{Ziqiang} \sur{Guan}}
\author[5]{\fnm{Jiemin} \sur{Li}}
\author[6,7]{\fnm{Byungjune} \sur{Lee}}
\author[8]{\fnm{Yu} \sur{Wang}}
\author[8]{\fnm{Seng Huat} \sur{Lee}}
\author[8]{\fnm{Zhiqiang} \sur{Mao}}
\author[9]{\fnm{Hoyoung} \sur{Jang}}
\author[5]{\fnm{Valentina} \sur{Bisogni}}
\author[5]{\fnm{Jonathan} \sur{Pelliciari}}
\author[4]{\fnm{Mark P. M.} \sur{Dean}}
\author[2,3]{\fnm{Steven} \sur{Johnston}}
\author[1]{\fnm{Matteo} \sur{Mitrano}}

\affil[1]{\orgdiv{Department of Physics}, \orgname{Harvard University}, \orgaddress{\city{Cambridge}, \state{MA}, \country{USA}}}

\affil[2]{\orgdiv{Department of Physics \& Astronomy}, \orgname{University of Tennessee}, \orgaddress{\city{Knoxville}, \state{TN}, \country{USA}}}

\affil[3]{\orgdiv{Institute of Advanced Materials and Manufacturing}, \orgname{University of Tennessee}, \orgaddress{\city{Knoxville}, \state{TN}, \country{USA}}}

\affil[4]{\orgdiv{Condensed Matter Physics and Materials Science Department}, \orgname{Brookhaven National Laboratory}, \orgaddress{\city{Upton}, \state{NY}, \country{USA}}}

\affil[5]{\orgdiv{National Synchrotron Light Source II}, \orgname{Brookhaven National Laboratory}, \orgaddress{\city{Upton}, \state{NY}, \country{USA}}}

\affil[6]{\orgdiv{Department of Physics}, \orgname{Pohang University of Science and Technology}, \orgaddress{\city{Pohang}, \country{Korea}}}

\affil[7]{\orgdiv{Max Planck POSTECH/Korea Research Initiative}, \orgname{Center for Complex Phase Materials}, \orgaddress{\city{Pohang}, \country{Korea}}}

\affil[8]{\orgname{2D Crystal Consortium}, \orgname{Pennsylvania State University}, \orgaddress{\city{University Park}, \state{PA}, \country{USA}}}

\affil[9]{\orgdiv{PAL-XFEL}, \orgname{Pohang Accelerator Laboratory, POSTECH}, \orgaddress{\city{Pohang}, \country{Korea}}}



\maketitle

\vspace{-10mm}
\section*{Supplemental Material}





\begin{enumerate}
\RaggedRight
\item \textbf{Hole doping and charge order}
\item \textbf{Analysis of RIXS spectra}
\item \textbf{DMRG calculations}

\end{enumerate}

\clearpage
\baselineskip24pt %

\section*{1. Hole doping and charge order}

The total fluorescence yield X-ray absorption spectroscopy and resonant soft X-ray scattering measurements were conducted at the RSXS endstation at the PAL-XFEL \cite{Jang2020time}. Samples were polished and mounted on a 6-axis in-vacuum manipulator. The incident X-rays were $\pi$-polarized and fluorescence and scattering signals were detected using avalanche photodiode point detectors. We corrected the X-ray absorption spectra for self-absorption effects using the package \textit{Larch} \cite{newville2013larch}. The algorithm employs a method developed by D. Haskell \cite{haskel1999fluo}, which applies an analytical correction based on the stoichiometry of the material and the incoming and outgoing x-ray angles. The magnitude of the effects was found to be of order 2\%.

Sr$_{14}$Cu$_{24}$O$_{41}$ is composed of Cu$_{2}$O$_{3}$ two-legged ladder and CuO$_{2}$ chain structural subunits separated by Sr layers. The ladders and chains have identical lattice parameters along the $a$ and $b$ direction, but are incommensurate along the $c$ axis ($c_L$ = 3.92 \AA, $c_C$ = 2.73 \AA). Assuming formal charges of Sr$^{2+}$ and O$^{2-}$, Sr$_{14}$Cu$_{24}$O$_{41}$ has a nominal average Cu valence of +2.25, with the excess holes shared between Cu atoms in the ladder and chain subunits\cite{osafune1997optical}. The hole density on the ladder is reported to be 0.06 holes/Cu, as quantified via O $K$-edge X-ray absorption spectra in previous studies \cite{nucker2000hole}. In Fig.~S1, we plot the normalized X-ray absorption spectra measured on our samples, together with previous measurements by Nucker, \textit{et al.}  \cite{nucker2000hole}. The datasets are in excellent agreement, particularly at the chain and ladder mobile carrier pre-peaks which encode information about the hole distribution. This confirms that the hole density of the ladder sublattice in our samples is 0.06 holes/Cu.

Our resonant soft X-ray scattering measurements show a sharp charge order peak resonant with the ladder mobile carrier pre-peak at 528.6 eV, at a momentum transfer (0, 0, 0.2 $\pm$ 0.014) r.~l.~u. (see Fig.~S2). We note that this momentum transfer is accessed using a different geometry than that used for the RIXS measurements. The charge order intensity decreases with increasing temperature, and vanishes at $T$\textsubscript{CO} = 255 K. The peak position and FWHM are independent of temperature. Our results are consistent with previous measurements \cite{abbamonte2004crystallization}, further confirming ladder hole density and the high quality of our single crystals. 

\section*{2. Analysis of RIXS spectra}

\subsection*{RIXS intensity due to quasiparticle spin flips}

In Fig. 2b of the main text, we highlight the contribution of quasiparticle (QP) spin flips to the RIXS intensity at 260 K, and at one representative momentum. Here, we describe the fitting procedure and show fits to spectra at additional representative momenta, measured at both 40 K and 260 K. The RIXS spectra below 0.7 eV exhibit several spectral features, including an elastic line, a phonon, its second harmonic, and a broad two-triplon continuum. At 260 K, an additional spectral feature, originating from QP spin flips, is evident at the low energy shoulder of the two-triplon continuum, particularly near the zone boundary. This feature closely follows the downward dispersion of the two-triplon continuum, and is only present above $T_\mathrm{CO}$, thus ruling out its assignment as a higher harmonic of the phonon. We fit the elastic, phonon, and QP spin flip features to Gaussian lineshapes. The two-triplon continuum is not a single mode, but its shape can be reproduced phenomenologically by the sum of an asymmetric Lorentzian and a Gaussian. We show the fit results at four representative momenta, measured at 40 K and 260 K, in Fig.~S5. The QP spin flip feature, highlighted in dark grey, is small or absent in the 40 K spectra (Fig.~S5(a-d)). In contrast, the QP spin flip peak is prominent in the 260 K spectra (Fig.~S5(e-h)), and disperses together with the low energy shoulder of the two-triplon continuum, consistent with theoretical predictions \cite{troyer1996properties, Kumar2019ladders}.

\subsection*{Extracting $S(\bm{q}, \omega)$ from the RIXS intensity}

The first step to convert the raw \gls*{RIXS} spectra (see Figs. S3 and S4) into dynamical spin structure factors is to subtract the non-magnetic components of the scattered signal. As shown in Fig.~2(b) of the main text, we fit to Gaussian lineshapes and subtract the elastic line, the 60-meV phonon, and its second harmonic at 120 meV.

Next, we normalize the subtracted \gls*{RIXS} spectra by scaling them by a geometry-dependent factor to ensure that the integrated intensity of the orbital excitations matches the theoretical single-ion scattering cross-section calculated using exact diagonalization (ED) \cite{Wang2019EDRIXS}. For this, we consider pure orbital excitations of a Cu$^{2+}$ ion in a square-planar crystal field, neglecting spin-flip and phonon contributions. In these calculations, $x$, $y$, and $z$ are parallel to the crystallographic $a$, $c$, and $b$ directions, respectively. We fix the theoretical crystal field parameters to match the experimental orbital excitation energies (neglecting the small dispersion of the d$_{xz/yz}$ orbital, see Fig. 4), and obtain $D_q$ = 0.164 eV, $D_s$ = 0.42 eV, and $D_t$ = 0.19 eV. The experimental and calculated \gls*{RIXS} spectra as a function of incident angle $\theta$ are shown in Fig.~S7(a) and S7(b). We multiply the experimental spectra in Fig.~S7(a) by a $\theta$-dependent scaling factor to match the integrated orbital intensity of the calculated spectra, plotted as a function of $\theta$ in Fig.~S7(c). We refer to these as `normalized \gls*{RIXS} spectra'.

Finally, we extract the dynamical spin structure factor $S(\bm{q}, \omega)$. The intensity of \gls*{RIXS} magnetic excitations is proportional to $S(\bm{q}, \omega)$ multiplied by the atomic form factor of the single-ion spin-flip amplitude $R\textsubscript{spin}(\epsilon , \epsilon^\prime, \Omega_i)$, where $\epsilon$ and $\epsilon^\prime$ are the polarization of the initial and final photons, and the $\Omega_i$ is the excitation energy \cite{ament2009theoretical,Robarts2021dynamical}:

 $$I_\text{spin} \propto R\textsubscript{spin}(\epsilon , \epsilon', \Omega_i) \times S(\bm{q}, \omega).$$

While this formalism does not account for doping, it has been shown that this correspondence is also applicable for doped systems within the energy range of magnetic excitations \cite{jia2014persistent}. We calculate $R\textsubscript{spin}(\epsilon , \epsilon', \Omega_i)$ for a single hole in a d$_{x^2 - y^2}$ orbital with its spin oriented along the in-plane diagonal direction [1, 0, 1] and $\pi$ polarized incident X-rays, following the approach outlined in previous work \cite{ament2009theoretical, shen2022role}. The calculated spin-flip scattering cross-section as a function of $\theta$ is shown in Fig.~S7(d). We obtain $S(\bm{q}, \omega)$ (up to an overall scaling factor) by dividing the normalized \gls*{RIXS} spectra by $R$\textsubscript{spin}. 

\subsection*{Orbital dispersion and Kugel-Khomskii model}

We assign  the \gls*{RIXS} orbital excitation peaks (see representative example in Fig.~4(a)) by using our single-site ED calculations outlined in the previous section. To visualize the dispersion of these excitations, we normalize the spectra to the integrated intensity between 1.4 eV and 3 eV and plot the data as a momentum-energy map in Fig.~4(b). We observe a clear dispersion in the lower branch of the d$_{xz/yz}$ excitation. The d$_{xy}$ and d$_{3z^2-r^2}$ branches are dispersionless. We additionally note a weak, dispersing shoulder at an energy loss of around 2.3 eV that is apparent only near the zone boundary.

We focus on the lower branch of the d$_{xz/yz}$ excitation, which is intense, well separated from other peaks, and shows a clear dispersion. The momentum-dependent orbital excitation spectra are fit to three peaks, neglecting the weak shoulder at 2.3 eV: a skewed Gaussian lineshape for the d$_{xy}$ branch, and Gaussian lineshapes for the d$_{xz/yz}$ and d$_{3z^2 - r^2}$ branches.

The center of the d$_{xz/yz}$ peak extracted from the fits is plotted as a function of momentum in Fig.~4(c), showing a dispersion of around 80 meV. The particular shape of the dispersion, with a maximum at the zone center and minima at approximately $\pm \pi/2$ is consistent with spin-orbital separation \cite{schlappa2012spin, wohlfeld2013microscopic, wohlfeld2019propagation}. Spin-orbital separation can be theoretically described using the Kugel-Khomskii model \cite{kugel1982jahn}. In this model, the lower branch of a propagating orbiton has a characteristic dispersion of the form $E(q) = E_0 - 2J_0\sin(q)$, where $E_0$ is the on-site energy of the d$_{xz/yx}$ orbital, $J_0$ is the orbital exchange constant, and $q$ is the momentum transfer along the $L$ direction. Fitting the dispersion of the d$_{xz/yz}$ excitation to this model, we obtain $E_0 = 2.013 \pm 0.007$ eV and $J_0 = 22.5 \pm 5.4$ meV (see Fig.~4(b)).

\clearpage

\section*{3. DMRG calculations}

\subsection*{Dynamical spin structure factors with $V$}

Fig.~S8 shows the $S(\bm{q},\omega)$ for various values of $V$ computed in a $64\times 2$ cluster. The model parameters are the same as above, with the modification that $U$ and $V$ are simultaneously changed to ensure $U-V=8t$. We find that positive (repulsive) $V$ and small values of negative (attractive) $V$ leave the spectra unchanged, with two-triplon excitations and an intense low-energy branch corresponding to the quasiparticle spin-flip excitations. However, further increasing the value of negative $V$ monotonically depletes the quasi-particle spin-flip scattering intensity. Larger negative values of $V$ also increase the paired holes tendency to collect near the boundaries of the cluster, consistent with previous results~\cite{zhou2023robust}. This behavior can be linked to their increased effective mass, which makes them more prone to localization. In this case, the boundary of the cluster creates a weak pinning potential, akin to an impurity. We have verified that any weak pinning potential in the cluster is sufficient to achieve this effect~\cite{Scheie2024}, including any incommensuration generated by the chains in Sr$_{14}$Cu$_{24}$O$_{41}$. 

\subsection*{Pair Correlations}
We examine the effect of $V$ on the superconducting correlations by calculating the rung-rung singlet pair correlation function 
\begin{equation}
    P_{r,r}(d) = \langle \Delta_r^\dagger(c)\Delta_r(c+d) \rangle, 
\end{equation} 
where the rung-singlet pair operator $\Delta_r(i) = \frac{1}{\sqrt{2}}(c_{i,1,\uparrow}c_{i,2,\downarrow} - c_{i,1,\downarrow}c_{i,2,\uparrow})$, as a function of distance from the center site $c$ of the cluster. Fig.~S9(a) plots the pair correlations at longer distances for representative values of $-|V|$. We find that $P_{r,r}(d)$ increases and begins to saturate at large distances as the attractive interaction is made stronger, consistent with the findings of reference~\cite{zhou2023robust}. These results indicate that the attractive next-nearest-neighbor interaction enhances the long-range superconducting correlations in the system. 

We also investigate the pairing symmetry of the ladder with $|V|\ne 0$ by comparing the rung-rung to the rung-leg pair correlations
\begin{equation}
    P_{r,l}(d) = \langle \Delta_r^\dagger(c)\Delta_l(c+d) \rangle,
\end{equation} 
where $\Delta_l(i) = \frac{1}{\sqrt{2}}(c_{i,1,\uparrow}c_{i+1,1,\downarrow} - c_{i,1,\downarrow}c_{i+1,1,\uparrow} )$ destroys the singlet pair along the leg of the ladder. When $|V| = 0$, $P_{r,r}$ and $P_{r,l}$ have the opposite sign, consistent with a $d$-wave-like pairing symmetry \cite{Nocera2018doping}. Our results show that this symmetry persists once $|V|\ne 0$, as shown in Fig.~S9(b).



\clearpage

\begin{figure}[h] 
    \centering
    \includegraphics[width=0.5\textwidth]{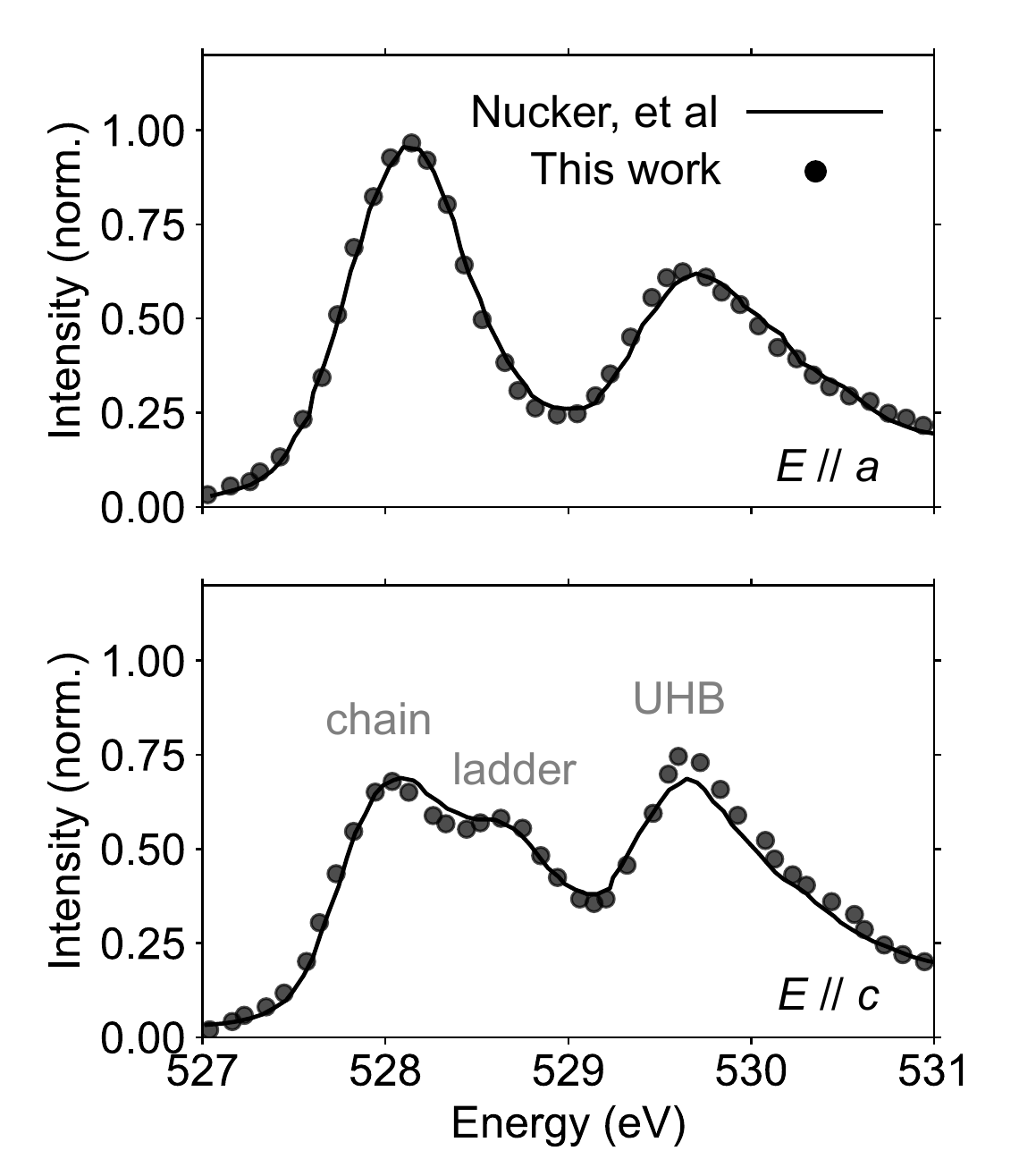} 
    \caption{{FIG.~S1. O $K$-edge X-ray absorption spectra.} Total fluorescence yield O K-edge X-ray absorption spectra measured on our samples and corrected for over-absorption (circles), overlaid on data measured by Nucker, et al \cite{nucker2000hole} (solid lines), for incident X-ray polarization along the $a$ (top) and $c$ (bottom) directions. The chain and ladder mobile carrier prepeaks and the upper Hubbard band (UHB) peak are labeled. The spectra are in excellent agreement with each other, confirming the hole density of 0.06 holes/Cu in our samples.}
\end{figure}

\clearpage

\begin{figure}[h] 
    \centering
    \includegraphics[width=\textwidth]{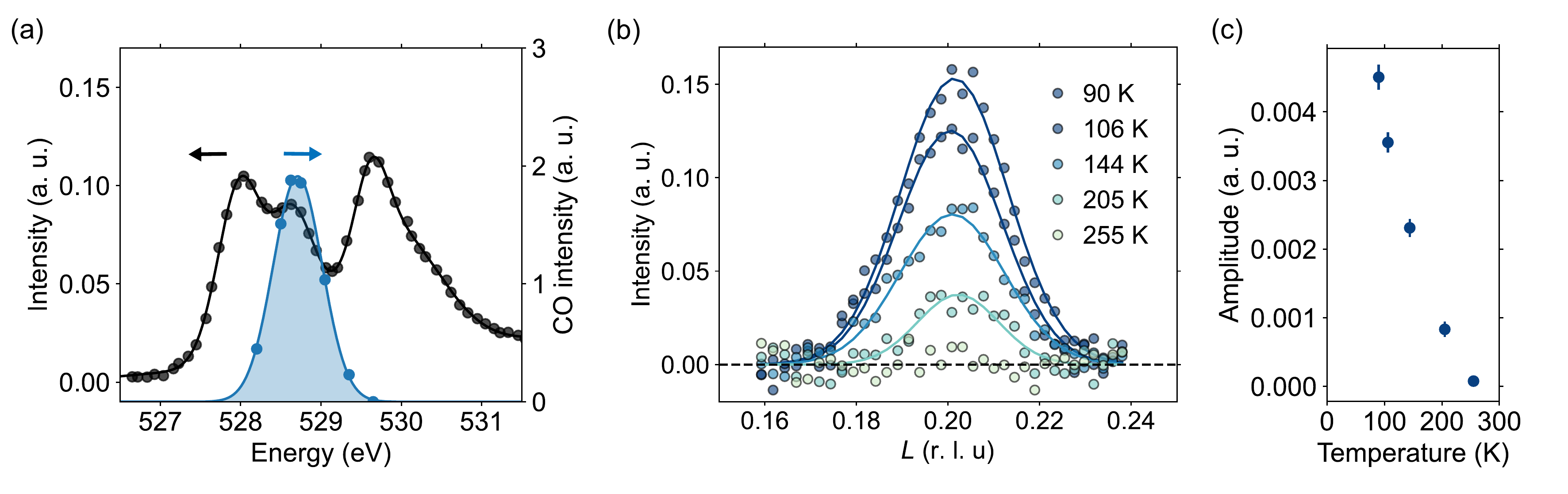} 
    \caption{{FIG.~S2. Charge order resonant with ladder holes.} (a) Charge order (CO) intensity as a function of incident X-ray photon energy (blue circles), overlaid on the O K-edge X-ray absorption spectrum with polarization along the $c$ direction (black circles). The CO peak is resonant with the ladder mobile carrier prepeak. (b) The CO diffraction peak at (0, 0, 0.20) r. l. u. at various temperatures. (c) The integrated intensity of the CO peak plotted as a function of temperature. The intensity vanishes at 255 K. The solid lines are fits to Gaussian lineshapes. r. l. u.: reciprocal lattice units}
\end{figure}

\clearpage

\begin{figure}[h] 
    \centering
    \includegraphics[width=1\textwidth]{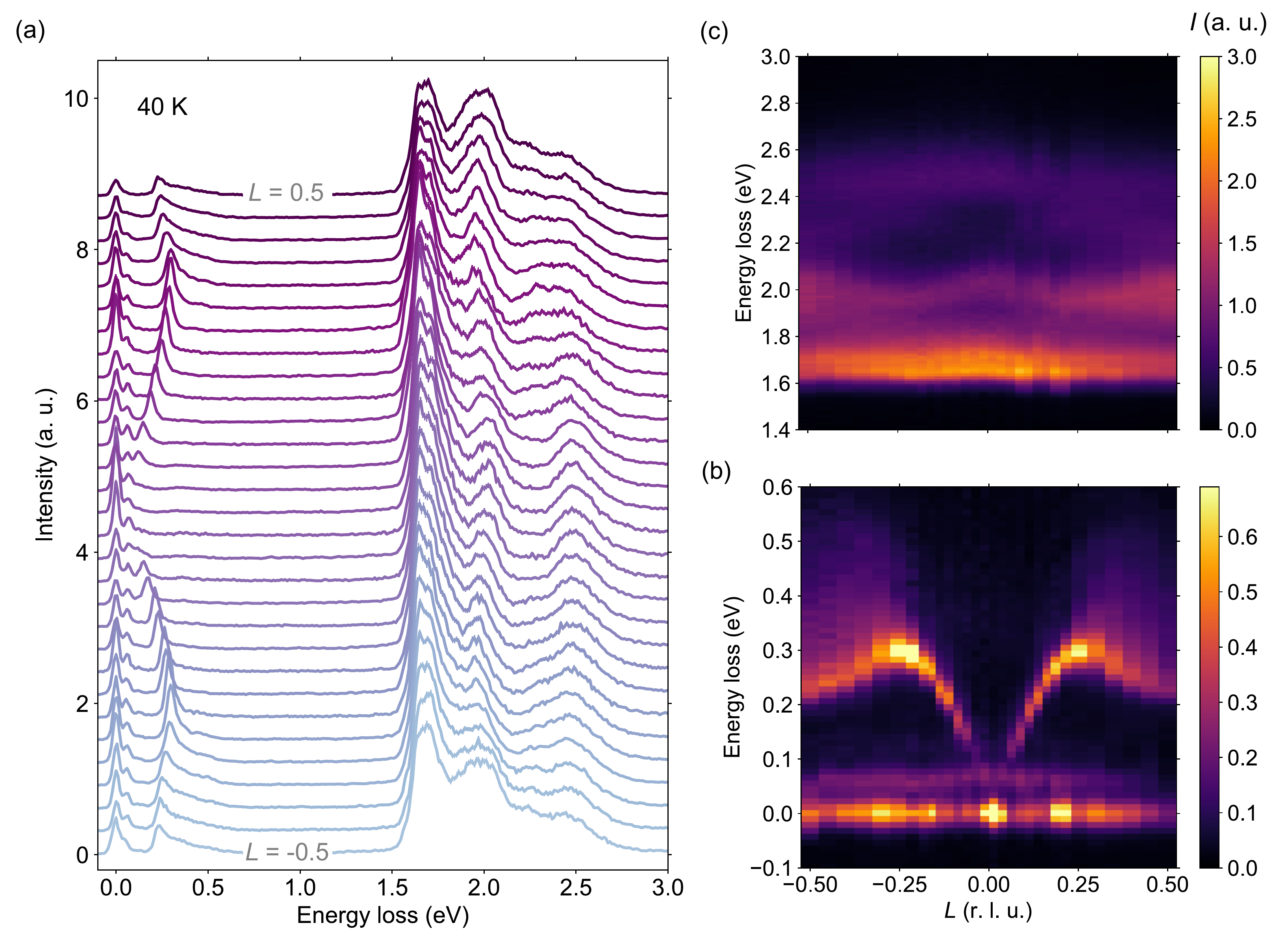} 
    \caption{{FIG.~S3. Raw \gls*{RIXS} spectra at 40 K.} (a) Raw \gls*{RIXS} spectra plotted as a function of momentum spanning $L$ = -0.5 to 0.5 r. l. u. measured at 40 K. (b-c) Momentum-energy maps of \gls*{RIXS} spectra in the regions dominated by magnetic excitations (b) and orbital excitations (c).}
\end{figure}

\clearpage

\begin{figure}[h] 
    \centering
    \includegraphics[width=1\textwidth]{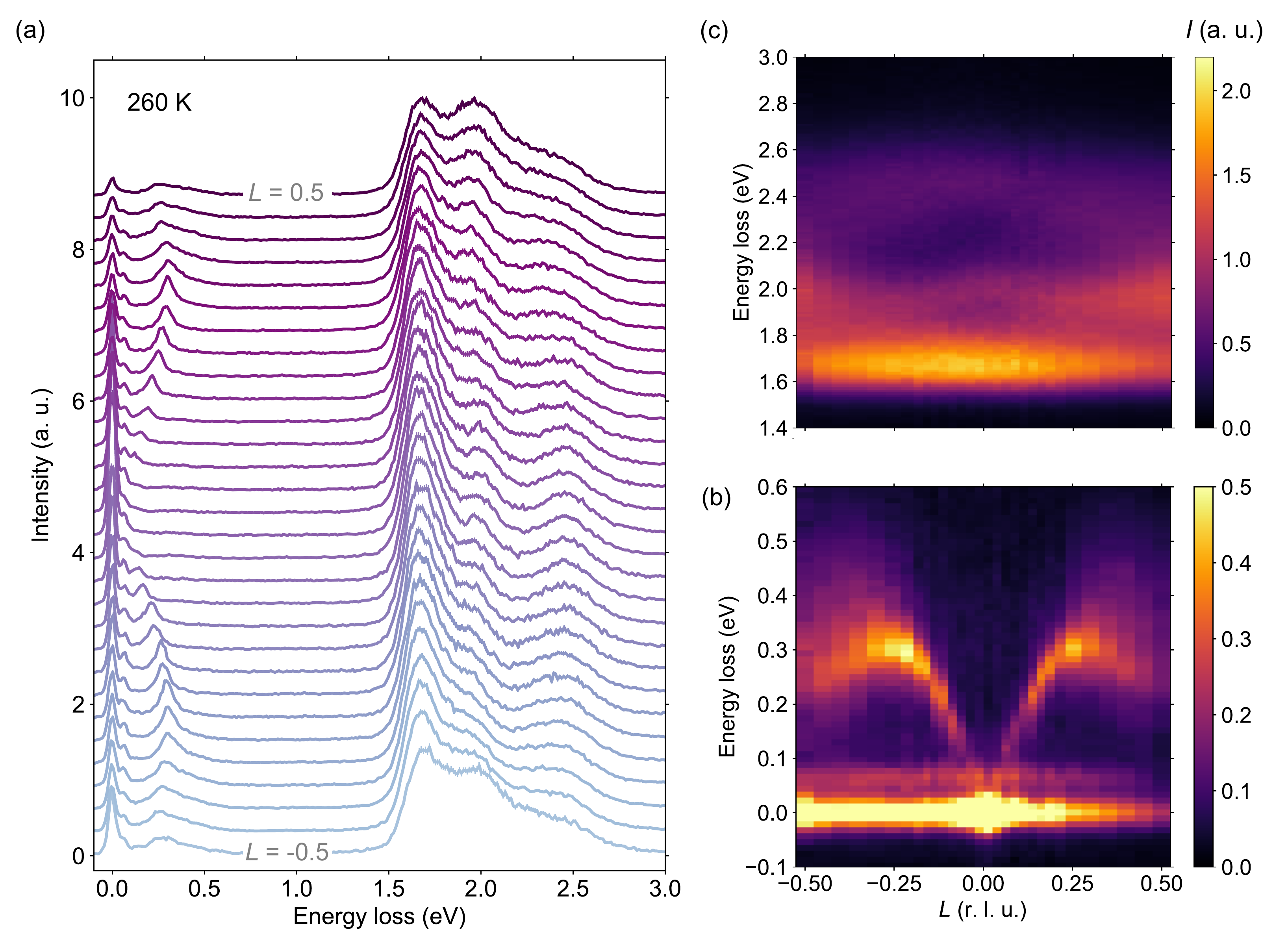} 
    \caption{{FIG.~S4. Raw \gls*{RIXS} spectra at 260 K.} (a) Raw \gls*{RIXS} spectra plotted as a function of momentum spanning $L$ = -0.5 to 0.5 r. l. u. measured at 260 K. (b-c) Momentum-energy maps of \gls*{RIXS} spectra in the regions dominated by magnetic excitations (b) and orbital excitations (c).}
\end{figure}

\clearpage

\begin{figure}[h] 
    \centering
    \includegraphics[width=1\textwidth]{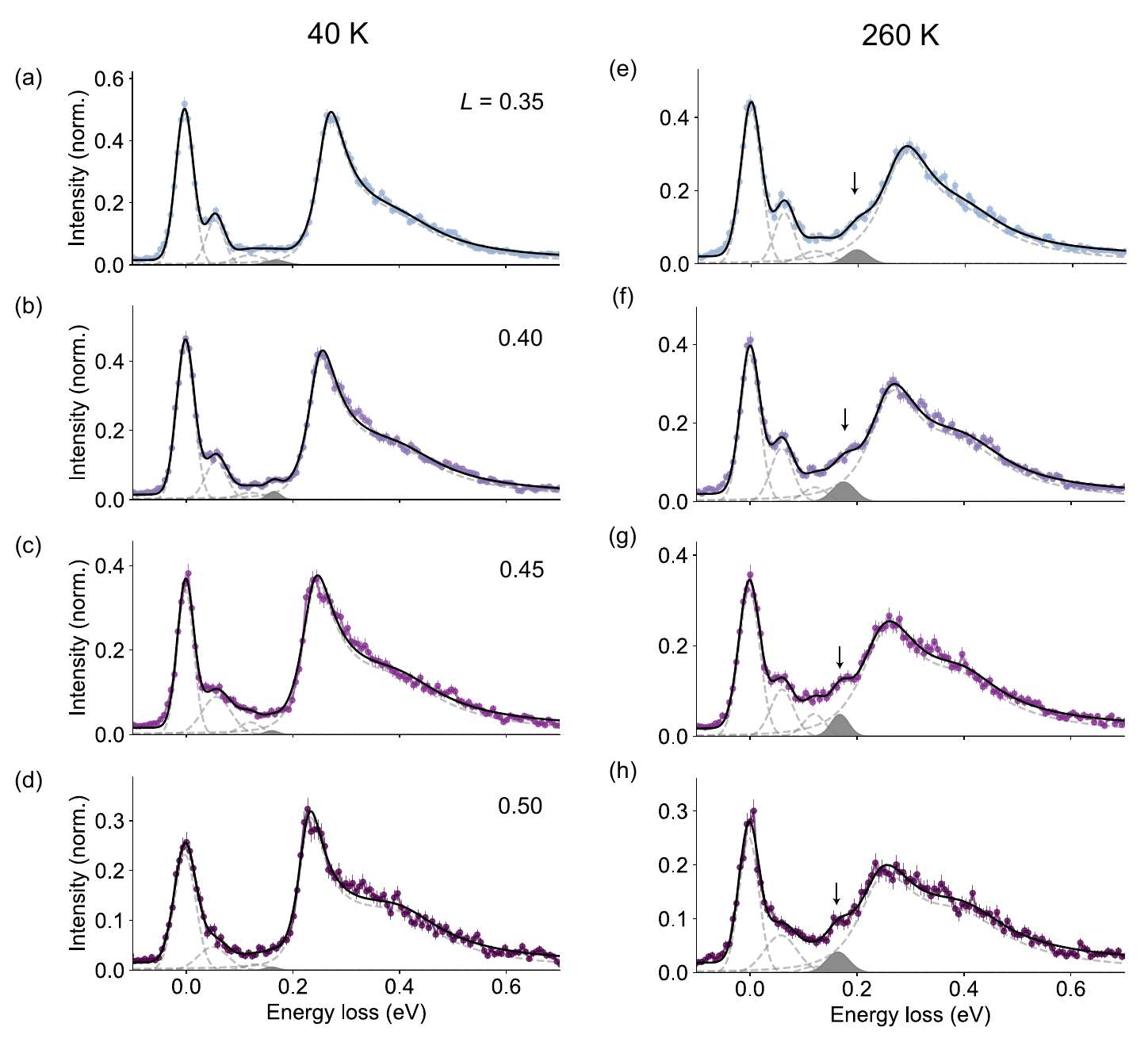} 
    \caption{{FIG.~S5. Quasiparticle spin flip contribution to RIXS intensity.} RIXS spectra measured at (a-d) 40 K and (e-h) 260 K, as a function of momentum $L$ = 0.35 to 0.50. The elastic, phonon, phonon harmonic, and quasiparticle spin flip features are fit to Gaussian lineshapes, and the two-triplon continuum is fit to the sum of an asymmetric Lorentzian and a Gaussian function. Black solid lines are the total fits, including a constant offset, grey dashed lines are individual contributions, and shaded dark grey regions are the quasiparticle spin flip contribution. The quasiparticle spin flip peaks are additionally indicated with arrows. }
\end{figure}

\clearpage

\begin{figure}[h] 
    \centering
    \includegraphics[width=0.6\textwidth]{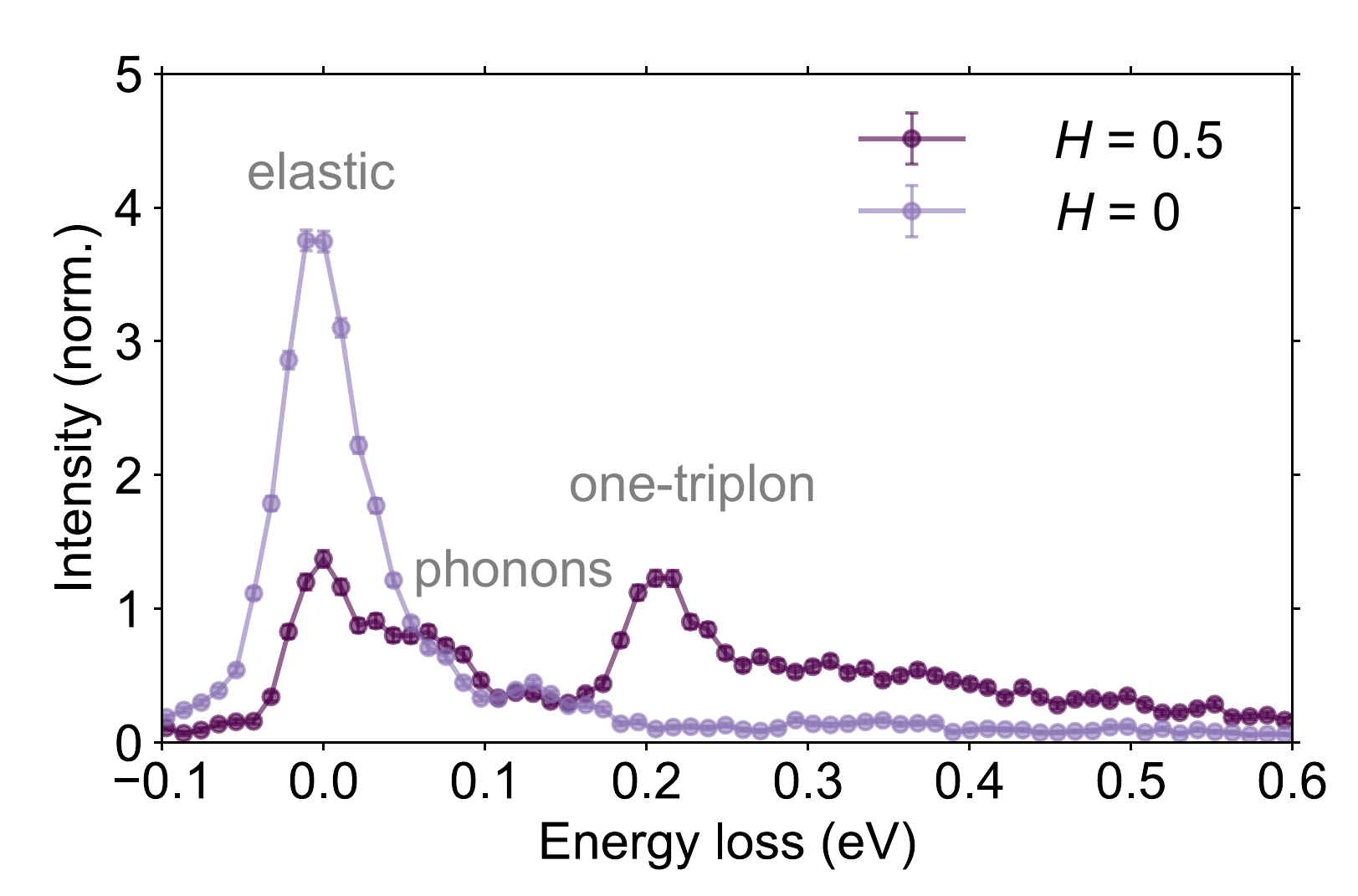} 
    \caption{{FIG.~S6. One-triplon excitation.} \gls*{RIXS} spectra measured at $H = 0.5$ and 0, with $L$ fixed to 0. The elastic, phonon, and one-triplon peaks are labeled. The spectra are normalized to the phonon intensity for comparison. The one-triplon excitation is observed only at $H = 0.5$, as expected from symmetry considerations.}
\end{figure}

\clearpage

\begin{figure}[h] 
    \centering
    \includegraphics[width=\textwidth]{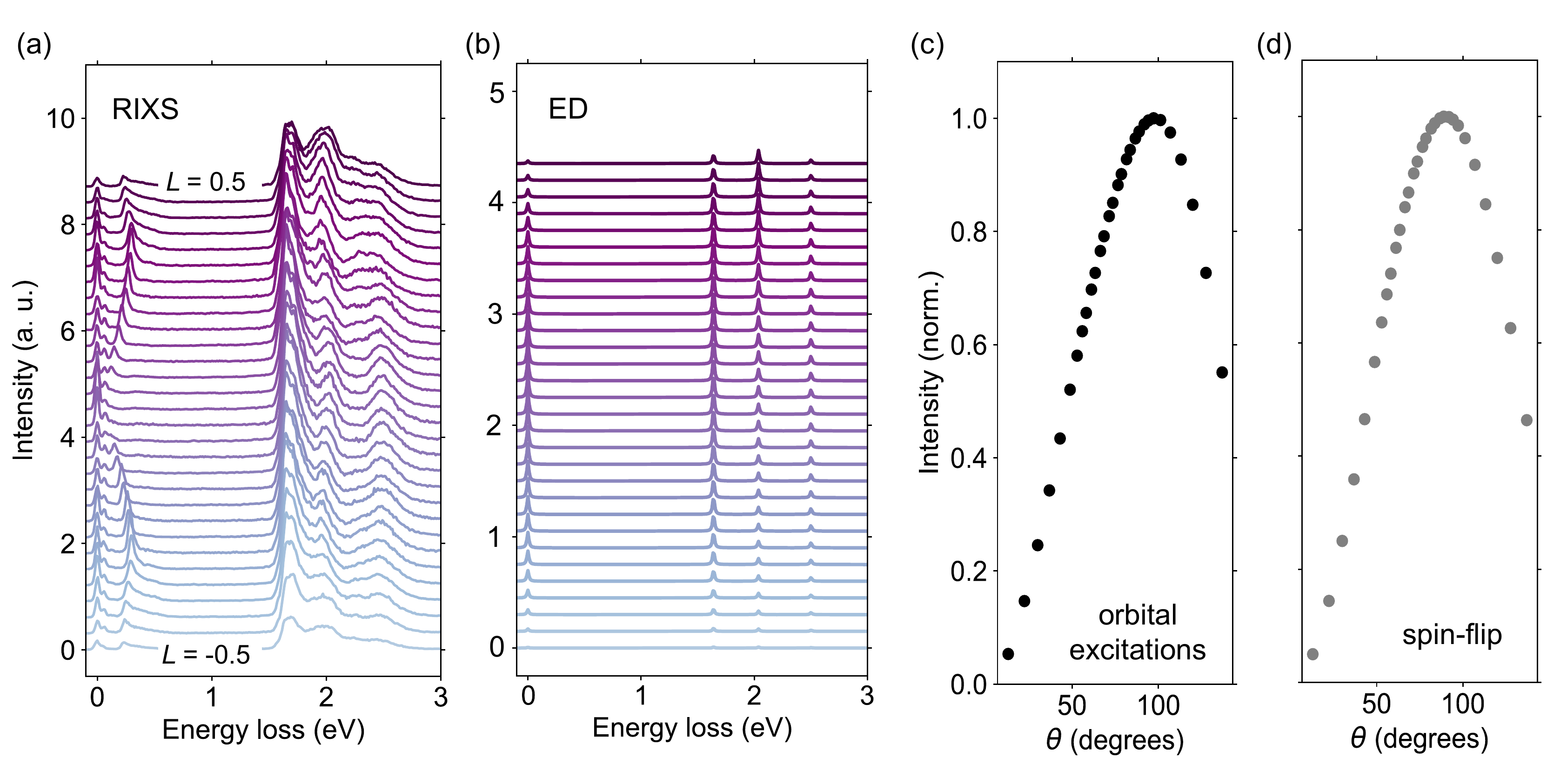} 
    \caption{{FIG.~S7. Single-site calculations using ED.} (a) The raw \gls*{RIXS} spectra measured as a function of momentum spanning $L$ = -0.5 to 0.5 r. l .u. (b) Orbital excitation spectra calculated using ED for a single-site model. (c-d) \gls*{RIXS} scattering cross-sections calculated using ED. Integrated orbital excitation scattering cross-section (c), and spin-flip cross-section (d) as a function of incident angle for $\pi$-polarized X-rays.}
\end{figure}

\clearpage

\begin{figure}[h] 
    \centering
    \includegraphics[width=\textwidth]{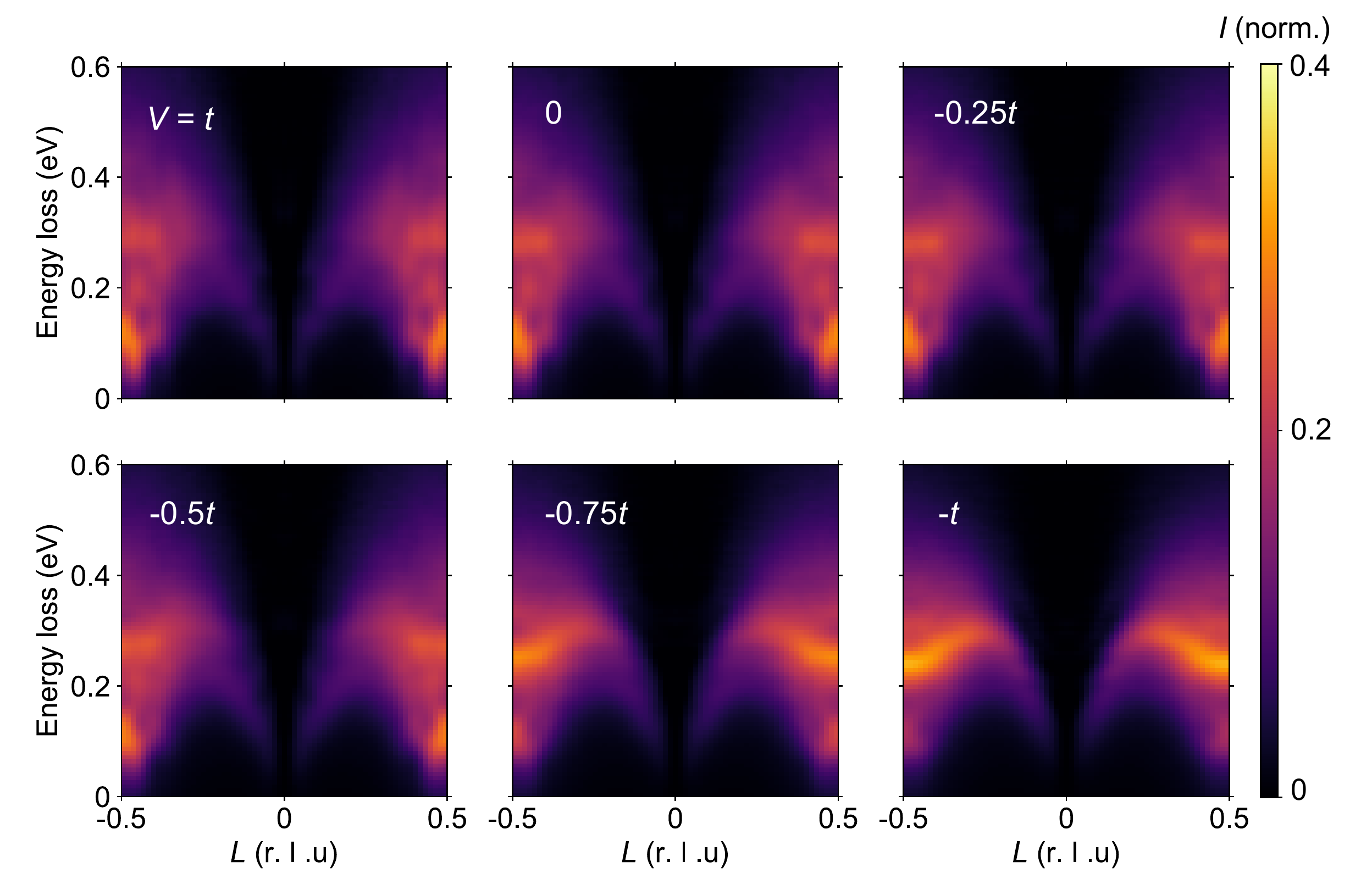} 
       \caption{{FIG.~S8. Dynamical spin structure factors for varying \textit{\textbf{V}}.} Density matrix renormalization group (DMRG) calculations for the dynamical spin structure factor computed on a $64\times 2$ cluster for the model parameters $t=1$, $t_\perp=0.84t$, $t^\prime=0.3t$, and $U - V = 8t$. $V$ ranges from $1$ to $-1$ and the broadening is set $\eta = 0.1t$. }
\end{figure}

\clearpage

\begin{figure}[h] 
    \centering
    \includegraphics[width=0.7\textwidth]{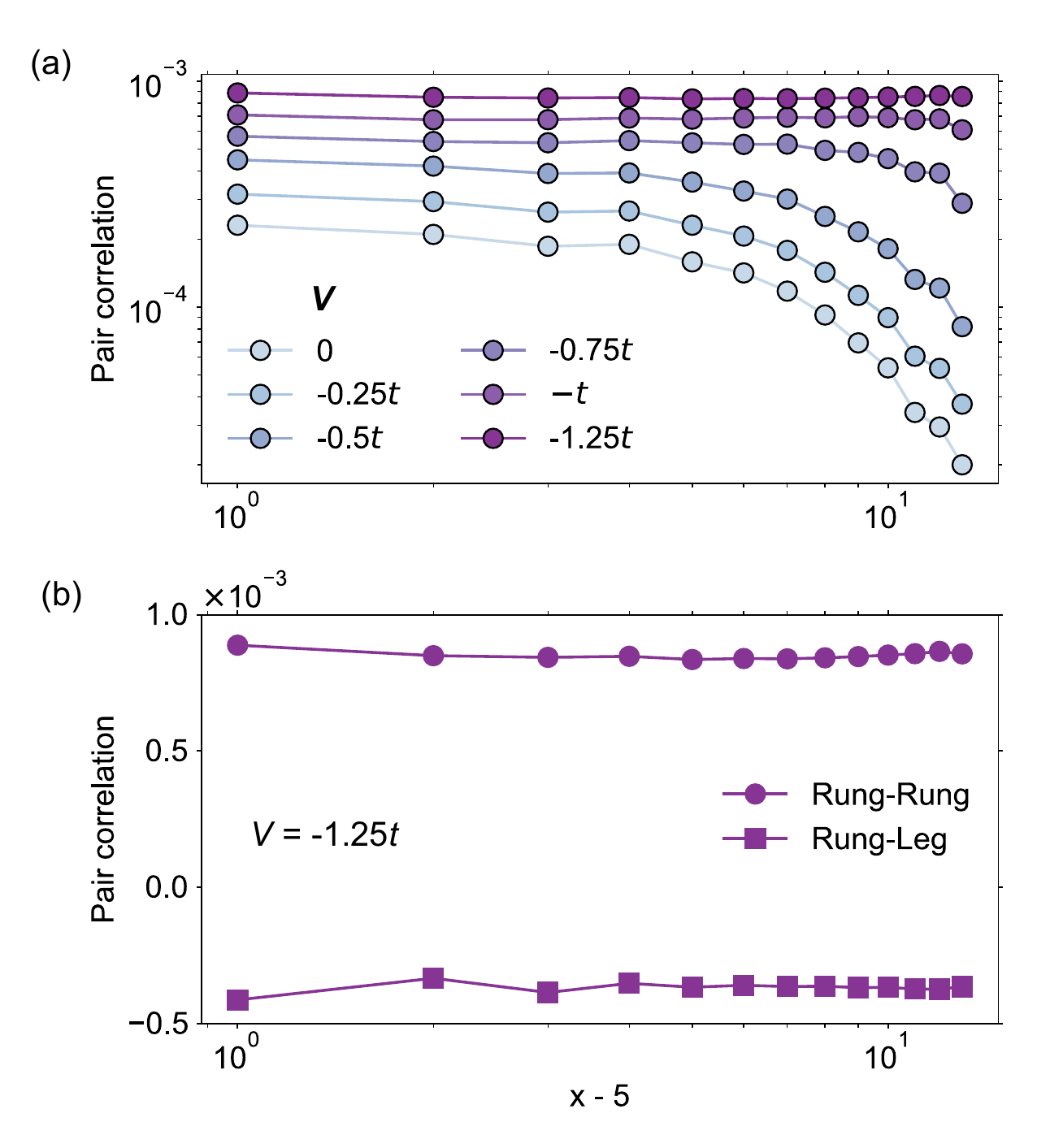} 
       \caption{{FIG.~S9. Singlet pair correlations.}~(a) DMRG results for the rung-rung singlet pair correlations for various values of $V$.~(b) Rung-rung and rung-leg singlet pair correlations for $V=-1.25t$. The sign change going from rung-rung to rung-leg directions indicates a $d$-wave-like character.}
\end{figure}

\clearpage
\bibliography{references}

%% file: glossary.tex
\newacronym{DMRG}{DMRG}{density matrix renormalization group}
\newacronym{RIXS}{RIXS}{resonant inelastic x-ray scattering}
\newacronym{XAS}{XAS}{x-ray absorption spectroscopy}
\newacronym{INS}{INS}{inelastic neutron scattering}
\newacronym{ARPES}{ARPES}{angle-resolved photo-emission}
\newacronym{CDW}{CDW}{charge-density wave}
\newacronym{2D}{2D}{two-dimensional}
\newacronym{1D}{1D}{one-dimensional}
\newacronym{BO}{BO}{bayesian optimization}